\documentclass[sigconf,nonacm]{acmart}
\setcopyright{none}
\renewcommand\footnotetextcopyrightpermission[1]{}
\usepackage{balance} % for balancing columns on the final page
\usepackage{amsmath}
\usepackage{subcaption}
\usepackage{algorithm}
\usepackage{algorithmic}
\usepackage{placeins}
\title[Network Topology Shapes AI Intervention Effectiveness on Deception]{Network Topology Shapes AI Intervention Effectiveness on Deception: Truth-Telling vs. Investigative Agents}

\author{Zhao Song}
\affiliation{\institution{Teesside University }\country{United Kingdom}}
\author{Stefan Sarkadi}
\affiliation{
  \institution{University of Lincoln}
  \country{United Kingdom}
}
\author{The Anh Han}
\affiliation{
  \institution{Teesside University}
  \country{United Kingdom}
}
\begin{abstract}
Deploying AI agents to combat dishonest communication is an intuitive response to deception in hybrid human-AI societies. However, the efficacy of such interventions in realistic networked settings remains unclear. We study this problem using an evolutionary game-theoretic model of the Deceiver's Game by extending it to structured networks, where agents can adopt deception, truth-telling, and investigation strategies. 
We introduce committed AI agents (zealots) fixed to either truth-telling or investigative strategies and conduct agent-based simulations across three network topologies. Our key findings are: (1) passive committed truth-telling agents can amplify adaptive deception by creating exploitable honest partners, whereas investigative agents increase adaptive truth-telling in the oscillatory regime; (2) intervention effects are non-monotonic in deployment density because committed agents reshape rather than simply suppress the underlying oscillatory dynamics; and (3) response peaks depend on topology, with sampled scale-free populations showing sharp low-density responses and fully connected populations requiring larger committed fractions.
Overall, the results highlight the need to distinguish passive truth-telling from active investigation when using committed agents to influence networked deception dynamics.
\end{abstract}


\newcommand{\BibTeX}{\rm B\kern-.05em{\sc i\kern-.025em b}\kern-.08em\TeX}

\begin{document}

%%% The following commands remove the headers in your paper. For final 
%%% papers, these will be inserted during the pagination process.

%\pagestyle{fancy}
%\fancyhead{}

%%% The next command prints the information defined in the preamble.

\maketitle 

%%%%%%%%%%%%%%%%%%%%%%%%%%%%%%%%%%%%%%%%%%%%%%%%%%%%%%%%%%%%%%%%%%%%%%%%

\section{Introduction}
Autonomous AI systems become increasingly capable of generating and disseminating deceptive information at scale, which poses a fundamental threat to trust in hybrid human-AI societies \cite{andras2018trusting,ferrario2020ai,jones2013design}. Recent empirical work demonstrates that AI-mediated communication can paradoxically increase dishonest behavior \cite{kobis2025delegation}, a phenomenon known as `moral offsetting' where humans offload ethical responsibility to AI systems. This raises an urgent question: How can we design AI interventions that actively counteract deception rather than amplify it?

One natural response is to deploy AI systems explicitly designed to promote truthfulness—AI agents that communicate honestly, refuse to deceive, and act as `truth-tellers' in multi-agent environments. However, the efficacy of such interventions is not yet well understood, particularly in realistic settings where agents interact through structured networks rather than in fully mixed populations. Do truth-promoting AI systems reliably suppress deception? Or can they introduce an effect similar to the `moral offset' and paradoxically backfire, reshaping population-level dishonesty dynamics in unintended ways?

Prior work on deception in evolutionary games has largely examined well-mixed populations, in which agents interact with one another with equal probability, see e.g. \cite{sarkadi2024triangles,mcnally2013cooperation,sarkadi2021evolution}. Yet many socio-technical systems are structured by relationships that shape who can communicate, exchange information, and influence behavior \cite{ravasz2003hierarchical,vespignani2012modelling}. Social media users interact through follower and friendship networks; organizations are organized into teams and reporting structures; and multi-agent AI systems may exchange information through defined communication links \cite{wooldridge1995intelligent,hammond2025multi}. Representing these settings as networks captures the fact that interactions are typically local rather than population-wide. This structure can alter the spread and persistence of strategies through network reciprocity \cite{nowak2006five,lamarins2024importance,szabo2007evolutionary,han2018fostering,bloembergen2015evolutionary,bara2022enabling,santos2006evolutionary}: cooperative agents (which are truth-tellers and investigators in our setting) may support one another in clusters, while deceivers may also benefit from clustering. Thus, network structure is not merely a feature of the interaction environment; it can shape the evolutionary dynamics of deception itself.

We hypothesize that topological structures shape whether AI-based interventions suppress or amplify deception. Specifically:

\textbf{H1 (Intervention Type):} Different types of AI interventions (truth-telling AI vs. investigative AI) produce opposite effects on population-level dishonesty.

\textbf{H2 (Non-monotonic Effects):} 
Truth-promoting or investigative AI interventions need not improve outcomes monotonically with deployment density. Intervention effects may peak at intermediate deployment densities, after which additional AI agents become less effective or counterproductive.
% Deploying more truth-promoting or investigative AI does not monotonically improve outcomes. An optimal deployment density exists beyond which additional AI agents become counterproductive.

\textbf{H3 (Network Topology):} The effect of AI deployment depends on the network structure in which agents interact. Scale-free networks, often used as stylized models of heterogeneous interaction structure, may respond differently than fully connected populations or regular lattices.
% The effect of AI deployment depends on the network structure in which agents interact. Scale-free networks (typical of social media and organizational hierarchies) respond differently than fully connected populations or regular lattices.

To address these questions, we study the evolution of dishonesty in networked populations using the Deceiver’s Game \cite{sarkadi2024triangles}, a framework for examining how deceptive and truth-telling strategies fare in social interactions. We extend this framework by introducing committed AI agents (zealots) \citep{masuda2012evolution,santos2019evolution, zimmaro2024emergence} that adhere consistently to either a truth-telling or an investigative strategy. This allows us to examine how fixed strategic commitments influence the spread of dishonesty across different network structures and how the role and prevalence of such agents may shape population-level outcomes.

Our findings suggest that intervention design should distinguish passive truth-telling from active verification. In this model, always-truthful agents can create exploitable partners for adaptive deceivers, whereas investigative agents change the payoff environment by challenging deception. The main implication is therefore mechanistic: intervention role, deployment density, and interaction structure jointly determine whether committed agents improve or worsen population-level truthfulness.
% Our findings suggest that practitioners designing trustworthy AI systems should approach truth-promoting interventions with caution. Deploying many AI agents trained to `always tell the truth' without considering network structure and interaction topology can backfire, inadvertently creating conditions that favor deception among adaptive agents. In contrast, AI systems designed to investigate, verify, and penalize dishonesty—even if deployed sparingly—provide a more reliable route to sustaining truthful communication across diverse network structures.
%%%%%%%%%%%%%%%%%%%%%%%%%%%%%%%%%%%%%%%%%%%%%%%%%%%%%%%%%%%%%%%%%%%%%%%%

The remainder of the paper is organized as follows. Section 2 reviews the relevant background and prior work. Section 3 presents our model and simulation methods. Section 4 reports the results, and Section 5 discusses their implications, limitations, and directions for future work. Additional methodological details and extensive robustness analyses are provided in the supplementary material.

\section{Related Work}
Research on deception in agent societies involves a significant literature, starting with the seminal prediction that AI agents will necessarily deceive us and each other \cite{castelfranchi2000artificial}, to the distinction between intentional and unintentional AI deception \cite{sarkadi2024deceptive}.

From the evolutionary multi-agent systems area, which is the most relevant to this work, we know several things. From \cite{sarkadi2024triangles}, we know that deception only becomes evolutionarily stable when agents care to do better than others, but not when agents care solely about their own payoffs. From \cite{sarkadi2024self}, we know that in order to re-establish resilient cooperation in the face of deception, agents need to account for the possibility of deception when calculating their trust in others that they haven’t interacted with before, i.e., they need to reduce their Truth-Default bias. However, this is cognitively demanding \cite{levine_truth-default_2014}.

Another crucial component of deception is that of Theory of Mind (ToM) \cite{sarkadi2019modelling}. That is the ability of agents to reason about the minds of other agents, usually in a recursive manner. The effect of ToM on deception is two-fold - it allows agents to outreason deceivers, but at the same time it creates pressure on deceivers to increase their meta-cognitive abilities \cite{sarkadi2023arms}.

The work of \cite{alon2026impomdp} shows that agents can actually mitigate the effect of deception when dealing with adversaries that have greater meta-cognitive abilities. However, as \cite{sarkadi2023arms} shows this can only be done in contexts 
where agents only care about their own payoffs. If instead, agents care to do better than others, then deception will become the optimal strategy. In other words, if the deceivers care to inflict deception on you at their own expense, then it is probably impossible to deter them.

Regarding the effect of deception on hybrid societies (human-AI teams), behavioral psychology reveals that delegation to AI can increase dishonest behavior \cite{kobis2025delegation}. This raises urgent questions about how to design trustworthy multi-agent environments \cite{paiva2018engineering,han2026socialphysicsageartificial}. Our work contributes to this challenge by showing that intervention type and network structure critically determine whether truth-oriented interventions succeed or backfire.

Finally, from the broader intervention literature on collective dynamics, we draw on research on committed minorities in opinion dynamics and evolution of prosocial behavior  \cite{becker2017network,lewandowsky2017beyond,santos2019evolution,song2026evolution,zimmaro2024emergence,perera2026hybrid}. These studies show that small committed groups can redirect population opinions \cite{galam2007role} and improve prosocial behaviors \cite{han2026socialphysicsageartificial}—but effects are highly context-dependent \cite{masuda2015opinion,mu2024multi}. We formalize this as deception games \cite{sarkadi2024triangles} on networks, showing that the effectiveness of committed agents depends on network topology and strategy type in non-monotonic ways. These findings offer design insights for self-organizing multi-agent systems, where committed agents may help shape collective behavior while accounting for the network structures and strategies that condition their influence.

\section{Model and Method}
We introduce the evolutionary game model, the networked agent-based simulation procedure, the committed-zealot intervention, and the selection-gradient method used to visualize macroscopic evolutionary dynamics. 

\subsection{Deceiver's Game X}

We build on Deceiver's Game X, introduced by~\citet{sarkadi2024triangles} to model dishonest communication in agent societies. We chose the respective model because it is able to capture, albeit at a higher level of abstraction, the socio-cognitive dynamics between deception and deception detection described in the related work section. The model contains three strategies,
\[S=\{D,T,I\},\]
where $D$ denotes deceivers, $T$ denotes truth-tellers, and $I$ denotes investigators. Deceivers strategically transmit information (false, true, or anything in between) that will make other agents infer false beliefs; truth-tellers communicate honestly without verification; investigators communicate honestly while paying a cost to verify others' claims.

When two agents use the same strategy, neither gains a relative advantage, and the payoff is normalized to zero. When a deceiver meets a truth-teller, the deceiver benefits from successfully exploiting an honest agent. This benefit is denoted by $\beta$, while $\theta$ denotes the intrinsic cost of deception. The deceiver therefore receives $\beta-\theta$, and the truth-teller receives $\theta-\beta$. When a truth-teller meets an investigator, both agents communicate truthfully, but only the investigator pays the verification cost $\gamma$. Thus the truth-teller receives $\gamma$, while the investigator receives $-\gamma$. When a deceiver meets an investigator, the investigator can detect or challenge deception. This interaction combines the deception cost $\theta$, the investigation cost $\gamma$, and an exposure penalty parameterized by $\epsilon$. The deceiver receives $\gamma-\theta-\epsilon(\gamma-\theta)$, while the investigator receives $\theta-\gamma+\epsilon(\gamma-\theta)$. Following the competitive formulation of Deceiver's Game X, payoffs represent relative strategic advantage rather than direct material utility.

Therefore, the payoff matrix is summarized as
\begin{equation}
M =
\begin{pmatrix}
0 & \beta-\theta & \gamma-\theta-\epsilon(\gamma-\theta) \\
\theta-\beta & 0 & \gamma \\
\theta-\gamma+\epsilon(\gamma-\theta) & -\gamma & 0
\end{pmatrix},
\label{eq:payoff-matrix}
\end{equation}
where rows and columns are ordered as $(D,T,I)$. 
Unless otherwise stated, we set $\beta=0.5$ and $\gamma=0.25$ following~\citet{sarkadi2024triangles}.

\subsection{Agent-based Simulations} 
A population of $N = 900$ agents occupies the nodes of a graph $G=(V,E)$. Each node hosts one agent, and edges determine possible interactions. For an agent $i$ with strategy $s_i$, the accumulated payoff is \[ P_i = \sum_{j\in \Omega_i} M_{s_i,s_j}, \] where $\Omega_i$ is the interaction neighborhood of $i$. Note that we use accumulated rather than degree-normalized payoffs, so payoff magnitudes vary with the number of interaction partners.

Strategies evolve through asynchronous Fermi imitation \cite{szabo2007evolutionary}. In each elementary update, an adaptive agent $i$ is selected uniformly at random and compared with a partner $j$ sampled from its interaction neighborhood. Agent $i$ adopts $j$'s strategy with probability
\[
W(s_i \leftarrow s_j)=
\frac{1}{1+\exp[(P_i-P_j)/K]},
\]
where $K$ quantifies the uncertainty in strategy adoption. We set $K=0.1$ throughout, as a typical choice in evolutionary game simulations on networks \cite{szabo2007evolutionary}. One Monte Carlo step (MCS) consists of $N$ elementary update attempts, so that each agent is selected once on average per MCS.

Unless otherwise stated, simulations reported in the main text include mutation with rate $\mu=10^{-5}$. Before imitation, an adaptive agent mutates to a randomly selected strategy with probability $\mu$. Mutation prevents finite-population absorption and allows the system to reveal long-run stochastic behavior rather than becoming trapped in a single absorbing state. Each stationary fraction reported in the simulation sweeps is averaged over 20 independent runs. Each run lasts $10^5$ MCS, and stationary fractions are computed over the final 5000 MCS. Algorithm~\ref{alg:simulation} summarizes the full simulation procedure.
\begin{algorithm}[htbp]
\caption{Agent-based simulation on networks.}
\label{alg:simulation}
\begin{algorithmic}[1]
\STATE Construct network $G=(V,E)$ with $|V|=N$
\STATE Initialize each agent's strategy $s_i \in \{D,T,I\}$ uniformly at random
\STATE Select zealot set $Z \subseteq V$ with fraction $z$
\STATE Fix $s_i=s^Z$ for all $i\in Z$
\FOR{$t=1$ to $T_{\max}$}
    \FOR{$n=1$ to $N$}
        \STATE Select agent $i$ uniformly at random
        \IF{$i\in Z$}
            \STATE continue
        \ENDIF
        \IF{random number $<\mu$}
            \STATE Set $s_i$ to a random strategy in $\{D,T,I\}$
        \ELSE
            \STATE Select comparison partner $j$ from $i$'s interaction neighborhood
            \STATE Compute payoffs $P_i$ and $P_j$ from Eq.~\eqref{eq:payoff-matrix}
            \STATE Agent $i$ adopts $s_j$ with probability
            $\left(1+\exp[(P_i-P_j)/K]\right)^{-1}$
        \ENDIF
    \ENDFOR
\ENDFOR
\STATE Return adaptive-agent strategy fractions over the observation window
\end{algorithmic}
\end{algorithm}
\subsection{Network Topologies}
We consider three interaction structures. The square lattice is a periodic two-dimensional lattice with degree four. The scale-free network is generated by a Barab\'asi--Albert preferential attachment process with average degree approximately four. The fully connected, or well-mixed, population is the complete graph, in which each agent interacts with all other $N-1$ agents. Additional checks on small-world and Erd\H{o}s--R\'enyi networks are reported in the supplementary material; these checks support the qualitative patterns emphasized in the main text.

These topologies allow us to compare local clustering, structural heterogeneity, and global well-mixed interaction \cite{cimpeanu2025evolutionary,santos2006evolutionary,szabo2007evolutionary}.

\begin{figure*}[t]
  \centering
  \begin{subfigure}[b]{0.85\textwidth}
    \includegraphics[width=\textwidth]{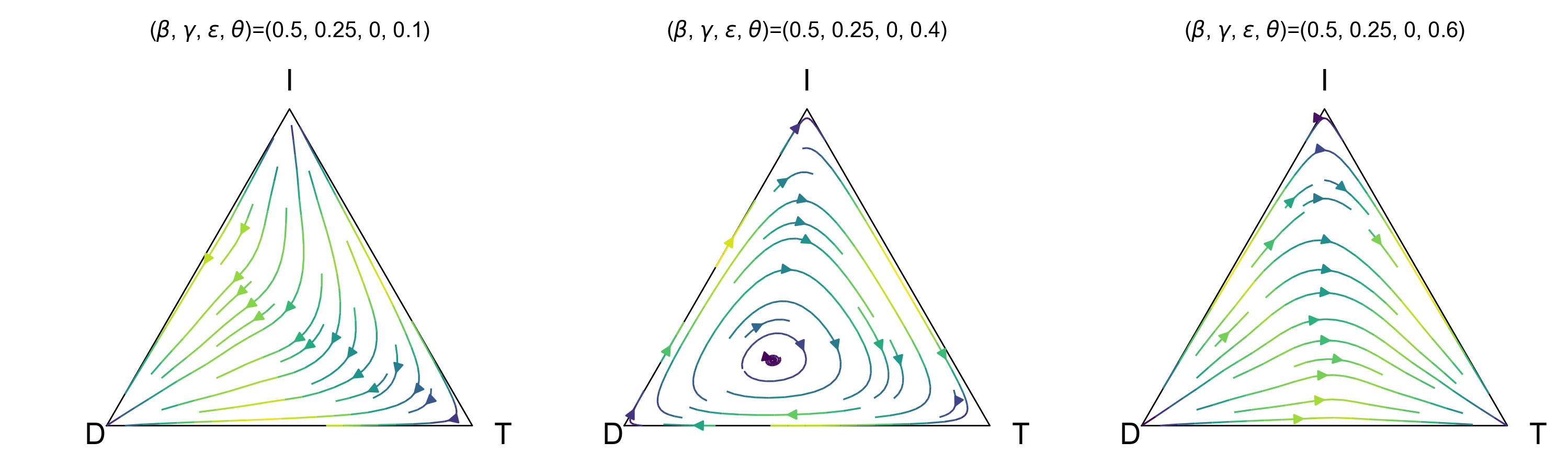}
  \end{subfigure}
  \hfill 
  \begin{subfigure}[b]{0.85\textwidth}
    \includegraphics[width=\textwidth]{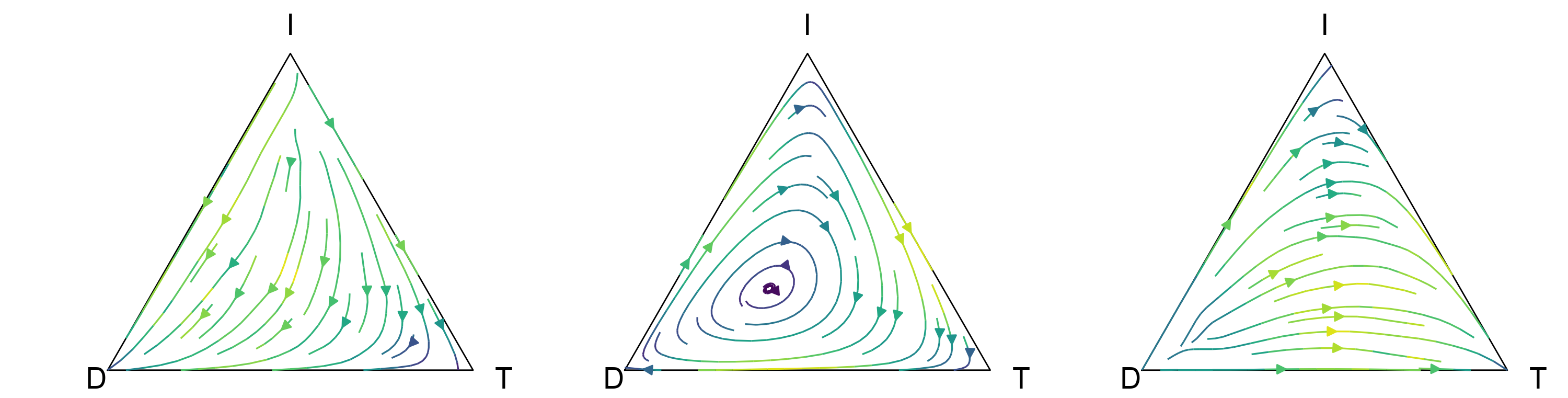}
  \end{subfigure}
  \hfill 
  \begin{subfigure}[b]{0.85\textwidth}
    \includegraphics[width=\textwidth]{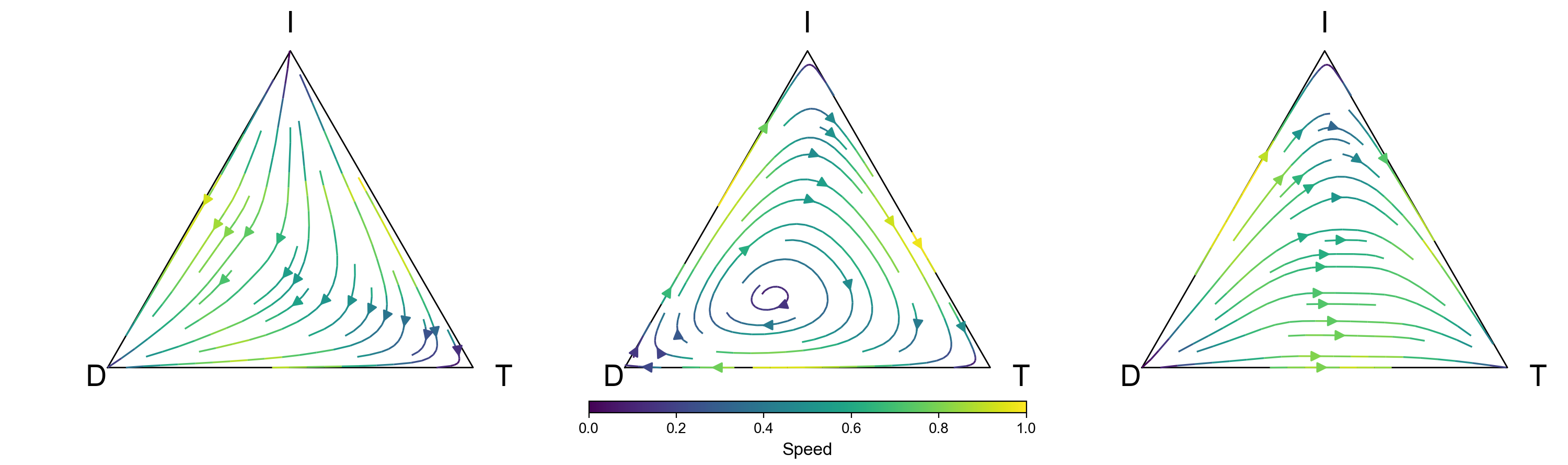}
  \end{subfigure}
  \caption{
  Networked populations contain an oscillatory regime.
  Rows show well-mixed, square lattice, and scale-free populations from top to bottom.
  Columns show increasing deception factor $\theta$ of 0.1, 0.4, and 0.6 under the no-penalty setting $\epsilon=0$.}
  \label{fig:1}
\Description{Three simplex selection-gradient plots for well-mixed, square lattice, and scale-free populations. Columns vary theta from 0.1 to 0.6 and show deception dominance, oscillation, and truth-telling regimes.}
\end{figure*}

\begin{figure*}[t]
  \centering
  \begin{subfigure}[b]{0.85\textwidth}
    \includegraphics[width=\linewidth]{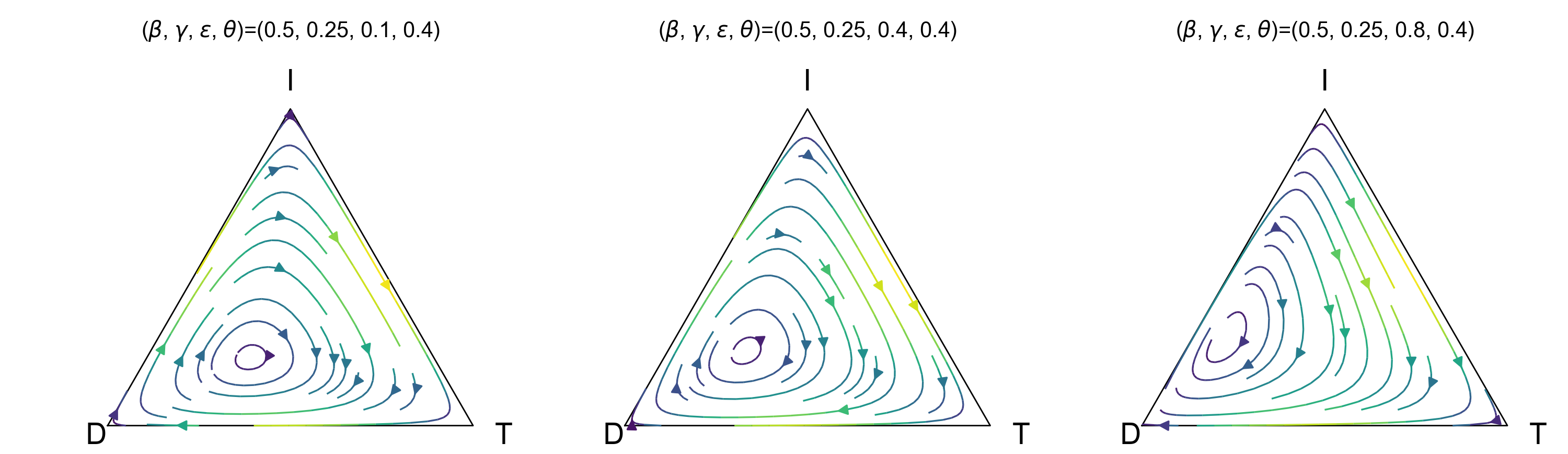}
  \end{subfigure}
  \hfill 
  \begin{subfigure}[b]{0.85\textwidth}
    \includegraphics[width=\linewidth]{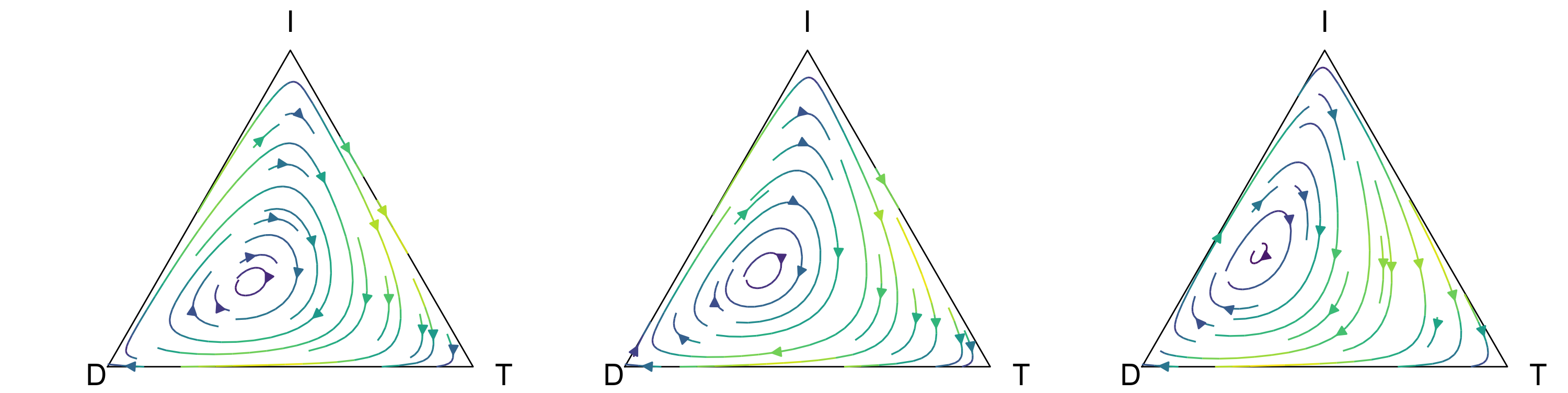}
  \end{subfigure}
  \hfill 
  \begin{subfigure}[b]{0.85\textwidth}
    \includegraphics[width=\linewidth]{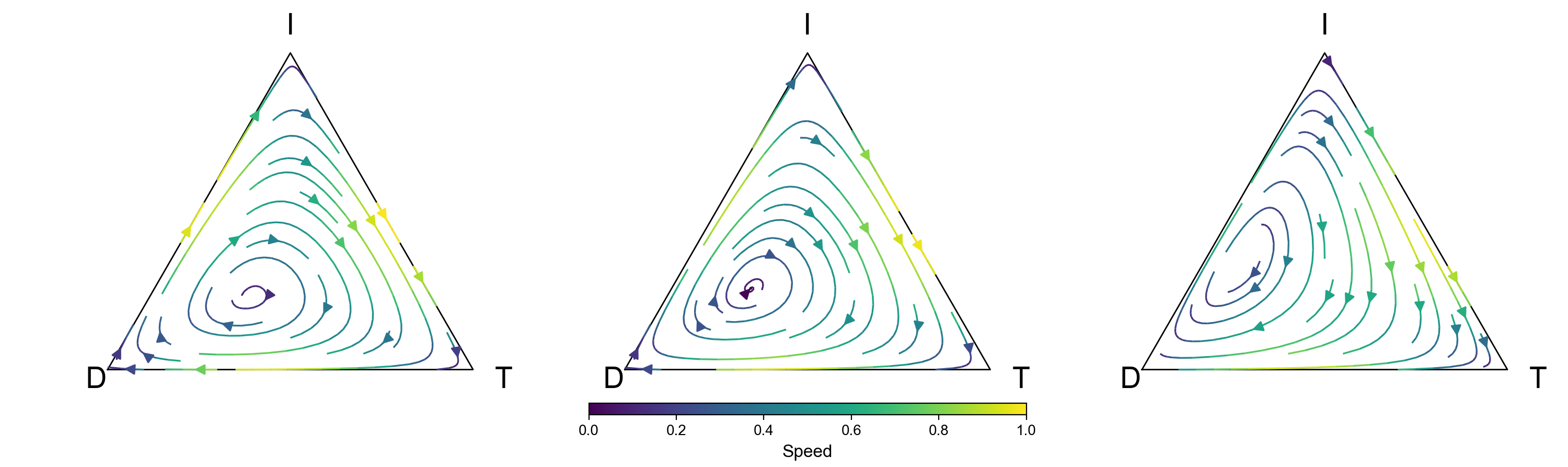}
  \end{subfigure}
  \caption{
  Effect of deception exposure penalty in the oscillatory regime.
  Rows show well-mixed, square lattice, and scale-free populations from top to bottom.
  Columns show exposure penalty values $\epsilon=0.1$, $\epsilon=0.4$, and $\epsilon=0.8$ at $\theta=0.4$.
  }
  \label{fig:3}
\Description{Selection-gradient plots for well-mixed, square lattice, and scale-free populations at theta 0.4 under increasing values of the exposure penalty epsilon.}
\end{figure*}

\subsection{Committed Zealots} 
To study intervention, we introduce committed zealots. A fraction $z$ of agents is selected uniformly at random and fixed to a prescribed strategy $s^Z\in\{D,T,I\}$. Zealots participate in payoff interactions and may be imitated by adaptive agents, but they never mutate or update their own strategy. In this paper, we focus on two cooperative intervention types: $T$-zealots, fixed as truth-tellers, and $I$-zealots, fixed as investigators. Unless otherwise stated, reported strategy fractions refer only to adaptive agents; the fixed zealot fraction is shown separately.
Under this mapping, $T$-zealots represent passive committed truth-tellers, whereas fact-checking or auditing roles are closer to $I$-zealots when they involve verification or challenge. 
In this abstraction, AI capability is represented behaviorally through commitment to a strategy, rather than through superior reasoning, memory, targeting, or verification accuracy.

\subsection{Selection-Gradient Visualization}
To visualize the macroscopic direction of evolution, we estimate the average selection gradient on the strategy simplex. Let $(\rho_D,\rho_T,\rho_I)$ denote the population fractions of the three strategies, with $\rho_D+\rho_T+\rho_I=1$.

For a given strategy configuration, the expected imitation-induced change in the frequency of strategy $a\in\{D,T,I\}$ over one MCS is
\[
\dot{\rho}_a =
\frac{1}{N}\sum_{i=1}^{N}\sum_{j\in\Omega_i}
\frac{1}{|\Omega_i|}
W(s_i\leftarrow s_j)
\left[\mathbf{1}(s_j=a)-\mathbf{1}(s_i=a)\right],
\]
where $\mathbf{1}(\cdot)$ denotes the indicator function, which equals $1$ if the condition inside the parentheses is satisfied, and $0$ otherwise.
We use $(\dot{\rho}_D,\dot{\rho}_T)$ as the vector field on the two-dimensional simplex, with $\rho_I=1-\rho_D-\rho_T$.
\begin{figure*}[htbp]
    \centering
    \includegraphics[width=0.95\linewidth]{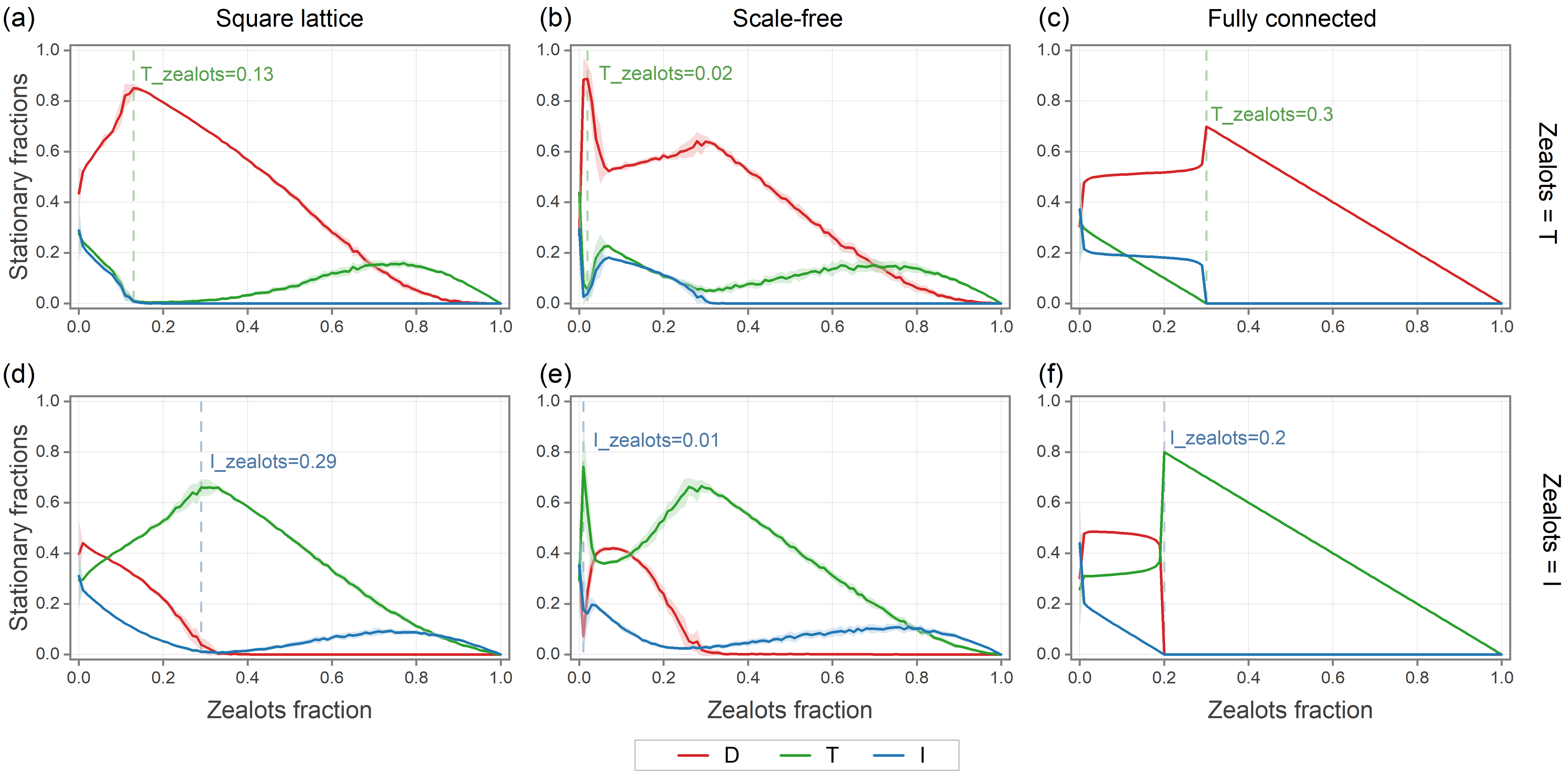}
    \caption{Committed truth-tellers and committed investigators produce distinct, non-monotonic effects across network topologies.
    Columns show square lattice, scale-free, and fully connected populations; rows show $T$-zealots and $I$-zealots.
    Solid curves show the stationary fractions of adaptive agents using $D$, $T$, and $I$; shaded bands show standard errors over 20 independent runs.
    Dashed vertical lines mark the zealot fractions that maximize adaptive deception for $T$-zealots and adaptive truth-telling for $I$-zealots.
    Parameters: $\theta=0.4$, $\epsilon=0$, $\mu=10^{-5}$.}
    \label{fig:zealots_topology}
\Description{Stationary adaptive strategy fractions under truth-teller and investigator zealots across lattice, scale-free, and fully connected networks. Curves show D, T, and I fractions with standard-error bands and peak-response markers.}
\end{figure*}

\begin{figure*}[h]
    \centering
    \includegraphics[width=0.95\linewidth]{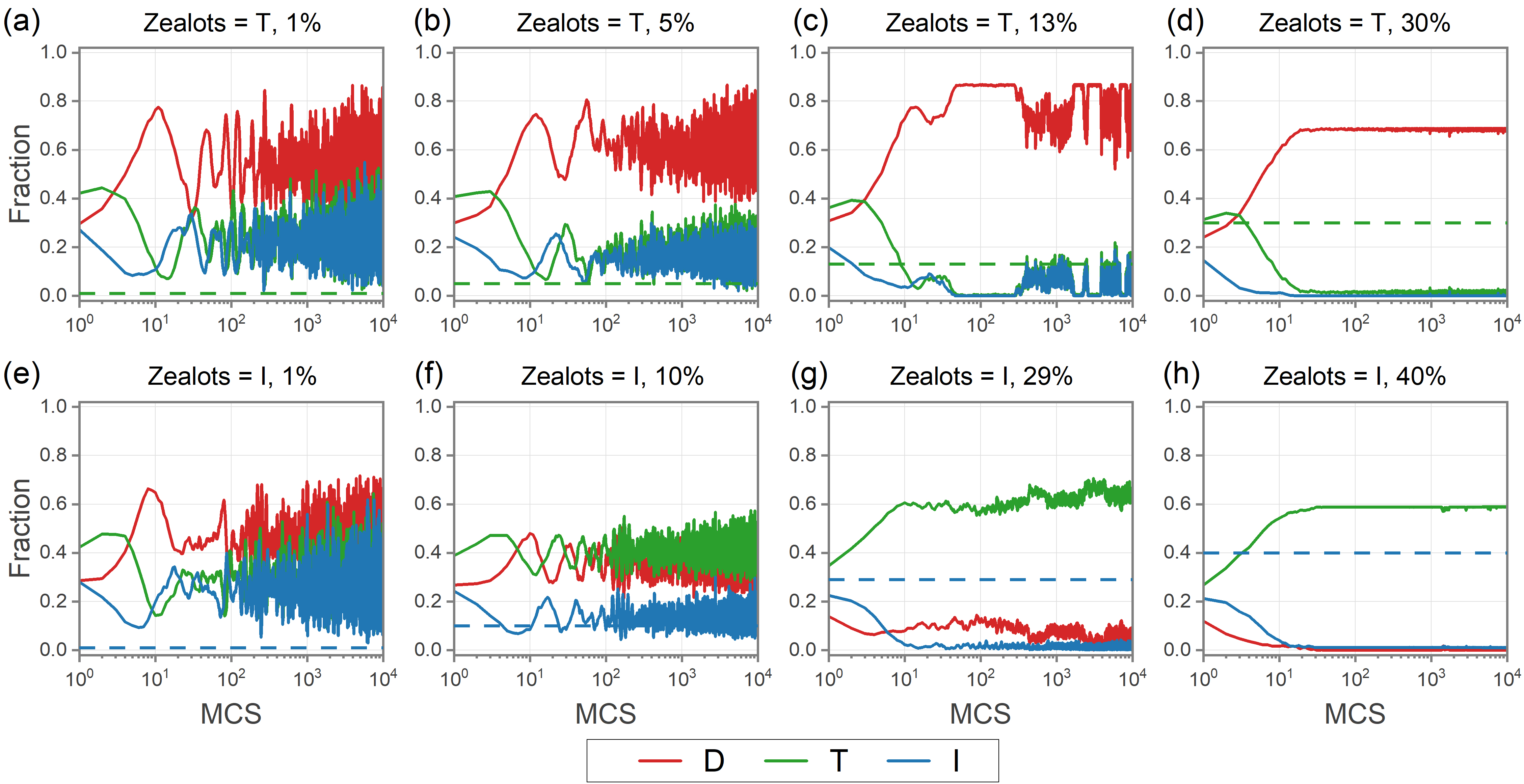}
    \caption{Selected Monte Carlo trajectories show how zealots reshape oscillations before suppressing them.
    Panels show square lattice time evolution for $T$-zealots at $1\%$, $5\%$, $13\%$, and $30\%$ in the top row, and for $I$-zealots at $1\%$, $10\%$, $29\%$, and $40\%$ in the bottom row.
    Solid curves show the fractions of adaptive agents using $D$, $T$, and $I$ over Monte Carlo steps; dashed horizontal lines indicate the fixed zealot fraction.
    Parameters: $\theta=0.4$, $\epsilon=0$, $\mu=10^{-5}$.
}
    \label{fig:zealots_MCS}
\Description{Monte Carlo trajectories for square lattice populations with different fractions of truth-teller and investigator zealots, showing how adaptive strategy fractions oscillate or become nearly stationary over time.}
\end{figure*}

\begin{figure*}[ht]
\centering
\includegraphics[width=0.95\linewidth]{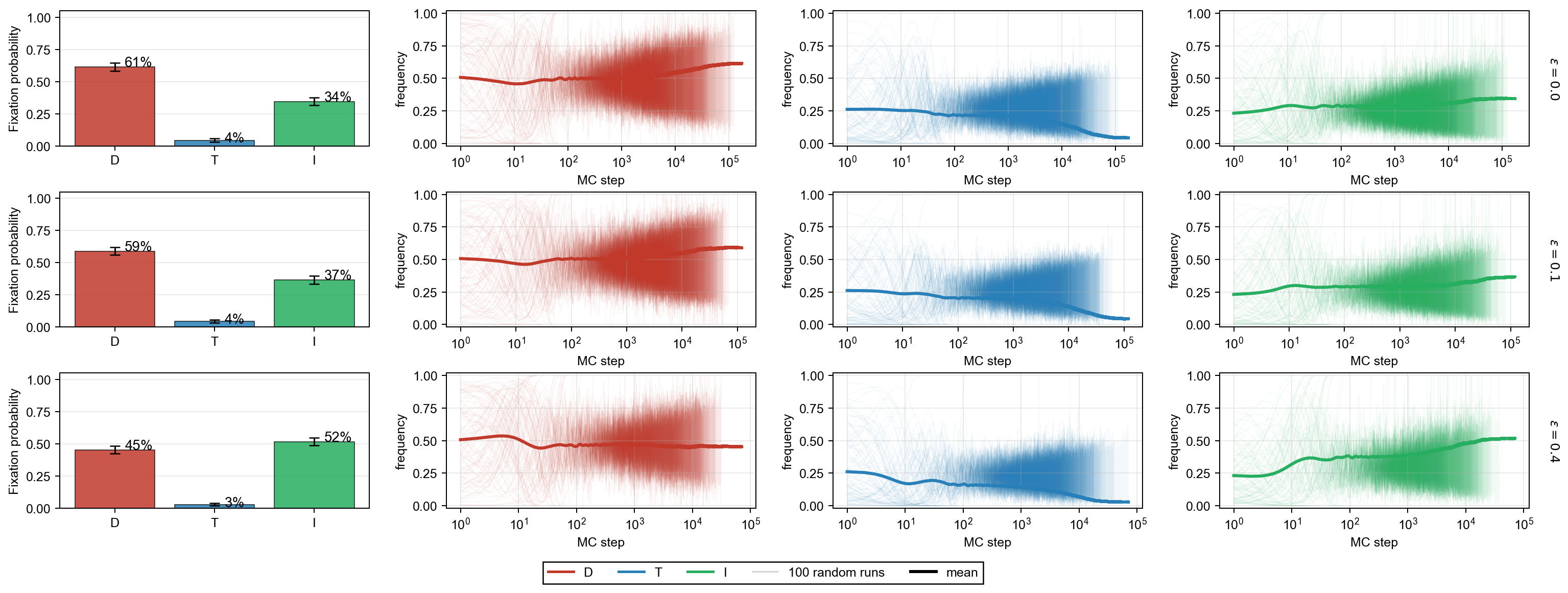}
\caption{
Repeated simulation check on the square lattice.
Panels show independent Monte Carlo simulations in the oscillatory regime at $\theta=0.4$ with $\epsilon=0$, $\epsilon=0.1$, and $\epsilon=0.4$.
Rows correspond to $\epsilon=0$, $\epsilon=0.1$, and $\epsilon=0.4$ from top to bottom.
The check is performed without mutation ($\mu=0$). The first column shows the probability of absorption into each homogeneous state across independent runs.
The remaining columns show the time evolution of the adaptive fractions of $D$, $T$, and $I$: faint curves are individual simulation trajectories and thick curves are averages across runs.
% Additional repeated-simulation checks are reported in the supplementary material.
}
\label{fig:lattice-repeated}
\end{figure*}
To estimate the average field in the selection-gradient plots, we enumerate initial compositions. For each initial composition, agents are randomly assigned strategies according to the specified fractions, and we perform 10 independent runs. During simulations, visited states are assigned to simplex bins, and the corresponding imitation-induced vectors are averaged over visits, initial conditions, and repetitions.

All trajectories used to estimate the displayed selection gradients include mutation with $\mu=10^{-5}$, but the displayed arrows report the imitation component only. Mutation is therefore used to avoid finite-population absorption during sampling, whereas the arrows represent payoff-based selection pressure.

\section{Results}

\paragraph{Deception dynamics oscillate on networks.} 
Networked populations do not generally converge to a single stationary strategy. Figure~\ref{fig:1} shows that the selection-gradient fields form rotating flows among deception, truth-telling, and investigation at intermediate deception factor $\theta=0.4$. This oscillatory regime appears across well-mixed, square lattice, and scale-free interaction structures, providing the dynamical setting in which committed agents are introduced below.

The oscillations raise a natural intervention question: whether a fixed minority of committed agents can redirect the adaptive population, and whether truth-telling is promoted most effectively by committed truth-tellers or by committed investigators.

\paragraph{Exposure penalties reshape, but do not eliminate, oscillations.}
% The oscillatory regime persists when the exposure penalty is introduced. 
As shown in Figure~\ref{fig:3}, the oscillatory regime persists when the exposure penalty is introduced.
At $\theta=0.4$, increasing $\epsilon$ changes the long-run composition of the cycle, shifting trajectories closer to the $D$--$I$ side of the simplex rather than stabilizing truth-telling. Thus, exposure penalties alone do not remove the cyclic tension among deception, truth-telling, and investigation; instead, they alter which strategies dominate the oscillatory regime.

\paragraph{Committed truth-tellers and committed investigators have opposite effects.}
The type of committed agent determines the direction of the population-level response. In the top row of Figure~\ref{fig:zealots_topology}, increasing the fraction of $T$-zealots raises the stationary fraction of adaptive deceivers over a broad range of zealot densities. In contrast, the bottom row shows that $I$-zealots suppress adaptive deception and increase adaptive truth-telling. Thus, committed truth-tellers can backfire by promoting deception among adaptive agents, whereas committed investigators enhance truth-telling.

\paragraph{Zealot effects peak at an intermediate density.}
The strongest effect is not obtained by adding as many zealots as possible. On the square lattice, $T$-zealots maximize adaptive deception at $z=0.13$, where adaptive deceivers occupy approximately $85.1\%$ of the non-zealot population; $I$-zealots maximize adaptive truth-telling at $z=0.29$, where adaptive truth-tellers occupy approximately $66.0\%$. In the fully connected population, the corresponding response peaks shift to $z=0.30$ for $T$-zealots and $z=0.20$ for $I$-zealots. The peak magnitudes also change, with fully connected interaction lowering the maximum adaptive deception induced by $T$-zealots to approximately $70.0\%$ and raising the maximum adaptive truth-telling induced by $I$-zealots to approximately $80.0\%$.

\paragraph{Scale-free networks are sensitive to very small committed minorities.}
Scale-free topology produces a distinct response at low zealot density. 
In Figure~\ref{fig:zealots_topology}(b), the strongest low-density response in the sampled scale-free populations occurs at $z=0.02$ for $T$-zealots, where adaptive deception reaches approximately $88.8\%$. 
In Figure~\ref{fig:zealots_topology}(e), the corresponding low-density response occurs at $z=0.01$ for $I$-zealots, where adaptive truth-telling reaches approximately $74.1\%$. 
After this low-density peak, the response declines and then rises again to a secondary high-density region around $z\approx0.29$--$0.30$. 
This creates a two-stage response in scale-free populations: a sharp low-density response peak followed by a weaker secondary response peak at intermediate zealot density.
% Scale-free topology produces a distinct response at low zealot density. In Figure~\ref{fig:zealots_topology}(b), only $2\%$ $T$-zealots are sufficient to maximize adaptive deception, which reaches approximately $88.8\%$. In Figure~\ref{fig:zealots_topology}(e), only $1\%$ $I$-zealots produce the largest adaptive truth-telling fraction, approximately $74.1\%$. After this low-density peak, the response declines and then rises again to a secondary high-density region around $z\approx0.29$--$0.30$. This creates a two-stage response in scale-free populations: a sharp low-density response peak followed by a weaker secondary response peak at intermediate zealot density.

\paragraph{Zealots reshape cycles before eliminating them.}
The non-monotonic stationary responses in Figure~\ref{fig:zealots_topology} arise from changes in the underlying oscillatory trajectories. To make this mechanism explicit, Figure~\ref{fig:zealots_MCS} focuses on the square lattice, where both zealot types exhibit clear intermediate response peaks in Figure~\ref{fig:zealots_topology}. At low zealot densities, Figures~\ref{fig:zealots_MCS}(a), (b), (e), and (f) show that the adaptive population continues to oscillate among the three strategies, even though the average composition is shifted. At the response peaks on the square lattice, Figures~\ref{fig:zealots_MCS}(c) and (g) show that the trajectories no longer explore the full three-strategy cycle: one adaptive strategy remains close to extinction for long periods, producing a high stationary fraction of adaptive deception under $T$-zealots and adaptive truth-telling under $I$-zealots. At larger zealot densities, Figures~\ref{fig:zealots_MCS}(d) and (h) show that additional adaptive strategies are excluded and the remaining adaptive population becomes nearly stationary. The peak response therefore occurs at the transition where oscillations are redirected into a reduced effective cycle, while excessive zealot density reduces the adaptive strategy space and weakens the target response.

\paragraph{Robustness and sensitivity analysis}
To test whether the oscillatory regime is an artifact of the small mutation rate used in the main simulations, we also ran repeated simulations without mutation. Figure~\ref{fig:lattice-repeated} shows the square lattice case at $\theta=0.4$ for $\epsilon=0$, $\epsilon=0.1$, and $\epsilon=0.4$. In the absence of mutation, independent runs typically exhibit transient cyclic dynamics before eventually reaching one of the absorbing homogeneous states. The first column reports the distribution of absorbing outcomes across runs, while the remaining columns show the individual trajectories and their mean. Thus, the small mutation rate used in the main simulations does not create the oscillatory tendency; rather, it prevents finite-population absorption and allows the underlying oscillatory regime to remain visible in the long-run dynamics. 

Additional checks are reported in the supplementary material. Additional repeated-simulation checks for small-world and well-mixed populations are shown in Supplementary Figure~B6. Mutation-rate sensitivity for the zealot experiments is shown in Supplementary Figure~B7. Further parameter and network checks, including Erd\H{o}s--R\'enyi and small-world networks, lattice-size checks, and broader parameter sweeps, are summarized in Supplementary Table~B1; representative results are shown in Supplementary Figures~B1--B5.

\section{Discussion}

\subsection{Why Do Truth-Tellers Backfire?}
A key mechanism in this work is that committed truth-tellers (T-zealots) amplify adaptive deception, whereas committed investigators (I-zealots) suppress it. One cause for this backfiring effect can be explained by the interaction between zealot dynamics and the underlying oscillatory game structure.

T-zealots are fixed under strategy updating, but their fixed truth-telling behavior makes them persistent exploitable partners for deceivers. From a deceiver's perspective, T-zealots are an ideal target: plentiful, honest, and non-retaliatory. At low T-zealot densities, adaptive deceivers can exploit this advantage, which makes deceivers increase in number. However, as T-zealot density rises, deceivers become so numerous that they begin to encounter each other and other adaptive agents more frequently, which then reduces the payoff advantage of deception relative to pure exploitation-free populations.

The strongest response occurs at intermediate density because this is where T-zealots maximally redirect the three-strategy cycle to favor deception without simultaneously suppressing adaptive variability. At this response peak, adaptive deceivers constitute approximately $85\%$ of the non-zealot population on the square lattice (Figure~\ref{fig:zealots_topology}(a)). Beyond this density, the adaptive strategy space collapses, and the population equilibrates near a stationary composition dominated by T-zealots and few adaptive deceivers—a situation that looks deceptively "bad" for truth-telling, but actually reflects population saturation rather than dynamic advantage.

In contrast, I-zealots succeed because they reshape the cycle differently. Investigators directly challenge and expose deceivers. Unlike truth-tellers, who are passive victims, investigators are active enforcers. At the I-zealot response peak, the cycle is redirected to keep deception suppressed and truth-telling elevated across a wide range of adaptive densities. Crucially, I-zealots do not suffer from truth-tellers' vulnerability: they are not exploitable, and they impose costs on deceivers.

In conclusion, passive interventions (truth-telling) can backfire, whereas active, enforcement-based interventions (investigation) are more resilient. 
This is consistent with arguments that truthful communication in the `post-truth era' requires active mobilization rather than passive information consumption~\cite{lewandowsky2017beyond}.
% This is in line with the argument that truthful communication can only be promoted in the `post-truth era' through active mobilization rather than passive information consumption \cite{lewandowsky2017beyond}.

\subsection{Real-World Implications for Hybrid Societies}
These findings provide model-based guidance for understanding how different AI intervention roles may affect dishonest communication in hybrid societies. 

First, our model distinguishes passive truth-telling from active verification. 
In our framework, $T$-zealots correspond to agents that consistently provide truthful information but do not verify or challenge others' claims, whereas fact-checking and auditing are closer to $I$-zealots. 
This distinction suggests that interventions based only on adding truthful content may be exploitable in adversarial communication environments, while verification-oriented agents can change the payoff environment faced by deceivers. 
This interpretation is consistent with empirical findings that downstream fact-checking is less effective than upstream source credibility assessment \cite{qi2025facts}.

Second, the results suggest that intervention density should be calibrated rather than maximized. 
Across the simulated networks, adding more committed agents does not always strengthen the desired response: intermediate densities can produce the strongest shift, whereas larger densities may saturate the adaptive strategy space. 
Thus, the model provides a mechanism for why truth-oriented interventions may have diminishing or even adverse returns when deployed without regard to population dynamics.

Finally, the results are relevant to autonomous multi-agent systems, where artificial agents increasingly interact, exchange information, and may generate dishonest communication \cite{castelfranchi2000artificial}. 
In such systems, simply adding agents committed to truthful statements may be less effective than introducing agents capable of verification and challenge. 
This interpretation resonates with debate-based approaches to truth discovery, where adversarial exchange and critical questioning are used to improve information quality \cite{sulkin2001habermas}. 
Future extensions should incorporate targeted placement, trust or credibility scores, imperfect verification, and heterogeneous AI capabilities.

\subsection{Non-Monotonicity and Intervention Mechanism}
A central contribution is showing that more committed agents are not always better. 
Zealot deployment produces non-monotonic responses, but the meaning of the peak depends on the intervention type: for $I$-zealots, it marks the highest adaptive truth-telling; for $T$-zealots, it marks the strongest backfiring response, where adaptive deception is maximized. 
Thus, these peaks should be interpreted as topology-dependent response points rather than universal intervention optima.

This non-monotonicity highlights three design lessons: intervention type matters, deployment density matters, and network structure matters. 
Investigator-type agents directly challenge deception, whereas truth-teller agents can provide exploitable honest partners; adding more committed agents can weaken the desired response once adaptive strategies become saturated; and scale-free populations can respond sharply to much smaller committed minorities than fully connected populations.

This contrasts with committed-minority models of opinion dynamics, where larger committed minorities typically strengthen the target opinion \cite{becker2017network}. 
In deception games, committed agents reshape payoff-mediated exploitation, so their effects can peak and then decline.

\subsection{Limitations and Future Work}
The main text focuses on three representative network structures, while supplementary checks include small-world and Erd\H{o}s--R\'enyi networks. Our model also assumes Fermi imitation, fixed payoff parameters, random zealot placement, and perfectly reliable investigation. Real socio-technical systems may involve heterogeneous preferences, trust- or credibility-based partner choice, targeted placement, costly communication and fallible verification \cite{bara2022enabling,han2021or,song2026network}. These extensions are important for future work, but the present model isolates a specific mechanism: intervention type, deployment density, and network structure jointly shape whether committed agents suppress or amplify deception.
% In this paper, we limited our experiments to representative network types. We assumed that agents' strategic responses match our model (Fermi imitation, structured networks) and that payoff parameters generalize across domains. Real socio-technical systems involve human psychology, heterogeneous preferences, and institutional constraints which we have not been able to capture here. However, we believe that the qualitative insight—that intervention type and network structure interact non-monotonically—should persist across many deception settings.

One future research direction is extending the model to temporal networks (where agents' relationships change over time, e.g., they enter and exit networks, or change their trust in each other, and also in regulators \cite{han2022voluntary}). Second, our zealots never mutate or update. In reality, we can also have partially committed agents, which in our context would represent institutions with dynamic AI regulation policies. Studying such partially committed agents might slightly improve realism, but is beyond the scope of this paper. Finally, empirical studies in human-AI hybrid teams, or lab experiments with humans and bot-agents, would validate whether the game-theoretic predictions align with real behavior.

\section{Conclusion}
In this paper, we have shown that embedding Deceiver's Game X in structured populations changes how committed AI roles affect dishonest communication. Passive truth-telling agents can increase adaptive deception by providing exploitable honest partners, whereas investigative agents promote adaptive truth-telling by challenging deception and changing the payoff environment faced by deceivers. These effects are non-monotonic in committed-agent density and depend on topology, including sharp low-density responses in sampled scale-free networks. This mechanism is consistent with prior work showing that decentralized peer-to-peer investigation and punishment can help re-establish cooperative knowledge sharing \cite{sarkadi2021evolution}, and with psychological evidence that deliberation improves resilience to false beliefs \cite{bago2020fake}. Taken together, the results provide model-based guidance for designing truth-oriented intervention roles in networked multi-agent systems: verification and challenge should be distinguished from simply adding truthful speakers. Future work should test the mechanism under targeted placement, imperfect verification, trust-based interactions, and empirical human-AI settings.
% In this paper, we show that network structure shapes the effectiveness of AI interventions against dishonest communication. Committed truth-telling AI agents, deployed naively, can paradoxically amplify deception by reshaping rather than suppressing oscillatory dynamics. Investigative AI agents (auditors, verification systems) are more robust, especially in sparse, heterogeneous networks. This aligns with the results in \cite{sarkadi2021evolution}, where only decentralized peer-to-peer investigation and punishment mechanisms were able to re-establish cooperative knowledge sharing, albeit only in populations smaller than 200 agents. It also aligns with experimental results in human fake news psychology that deliberation (slow thinking) is more resilient in the face of false beliefs \cite{bago2020fake}. Our findings have immediate implications for designing trustworthy multi-agent systems, from social media moderation to organizational governance to AI-AI coordination protocols. We argue that future work should test these predictions empirically on real agent networks, explore adversarial adaptation, and extend the model to capture various instantiations of the deception strategy and network topologies.

\balance
\bibliographystyle{ACM-Reference-Format}
\bibliography{mybib}

@article{song2026evolution,
  title={Evolution of fairness in hybrid populations with specialised AI agents},
  author={Song, Zhao and Cimpeanu, Theodor and Shen, Chen and Han, The Anh},
  journal={Philosophical Transactions of the Royal Society A. Preprint available arXiv:2602.18498},
  year={2026}
}

@article{han2026socialphysicsageartificial,
      title={Social physics in the age of artificial intelligence}, 
      author={The Anh Han and Andrea Baronchelli and Joel Z. Leibo and Tom Lenaerts and Iyad Rahwan and Fernando Santos and Matjaž Perc and Valerio Capraro},
      year={2026},
      journal = {Phil. T. Roy. Soc. A (in press). Preprint arxiv.org/abs/2603.16900},
      eprint={2603.16900},
      archivePrefix={arXiv},
      primaryClass={physics.soc-ph},
      url={https://arxiv.org/abs/2603.16900}, 
}

@article{andras2018trusting,
  title={Trusting intelligent machines: Deepening trust within socio-technical systems},
  author={Andras, Peter and Esterle, Lukas and Guckert, Michael and Han, The Anh and Lewis, Peter R and Milanovic, Kristina and Payne, Terry and Perret, Cedric and Pitt, Jeremy and Powers, Simon T and others},
  journal={IEEE Technology and Society Magazine},
  volume={37},
  number={4},
  pages={76--83},
  year={2018},
  publisher={IEEE}
}

@article{ferrario2020ai,
  title={In AI we trust incrementally: A multi-layer model of trust to analyze human-artificial intelligence interactions},
  author={Ferrario, Andrea and Loi, Michele and Vigan{\`o}, Eleonora},
  journal={Philosophy \& Technology},
  volume={33},
  number={3},
  pages={523--539},
  year={2020},
  publisher={Springer}
}

@article{jones2013design,
  title={The design of intelligent socio-technical systems},
  author={Jones, Andrew JI and Artikis, Alexander and Pitt, Jeremy},
  journal={Artificial Intelligence Review},
  volume={39},
  number={1},
  pages={5--20},
  year={2013},
  publisher={Springer}
}

@article{vespignani2012modelling,
  title={Modelling dynamical processes in complex socio-technical systems},
  author={Vespignani, Alessandro},
  journal={Nature physics},
  volume={8},
  number={1},
  pages={32--39},
  year={2012},
  publisher={Nature Publishing Group UK London}
}

@article{ravasz2003hierarchical,
  title={Hierarchical organization in complex networks},
  author={Ravasz, Erzs{\'e}bet and Barab{\'a}si, Albert-L{\'a}szl{\'o}},
  journal={Physical review E},
  volume={67},
  number={2},
  pages={026112},
  year={2003},
  publisher={APS}
}

@article{mcnally2013cooperation,
  title={Cooperation creates selection for tactical deception},
  author={McNally, Luke and Jackson, Andrew L},
  journal={Proceedings of the Royal Society B: Biological Sciences},
  volume={280},
  number={1762},
  pages={20130699},
  year={2013}
}

@inproceedings{han2018fostering,
  title={Fostering cooperation in structured populations through local and global interference strategies},
  author={Han, The Anh and Lynch, Simon and Tran-Thanh, Long and Santos, Francisco C and others},
  booktitle={27th International Joint Conference on Artificial Intelligence and the 23rd European Conference on Artificial Intelligence},
  pages={289--295},
  year={2018},
  organization={IJCAI}
}

@article{bloembergen2015evolutionary,
  title={Evolutionary dynamics of multi-agent learning: A survey},
  author={Bloembergen, Daan and Tuyls, Karl and Hennes, Daniel and Kaisers, Michael},
  journal={Journal of Artificial Intelligence Research},
  volume={53},
  pages={659--697},
  year={2015}
}

@article{santos2006evolutionary,
  title={Evolutionary dynamics of social dilemmas in structured heterogeneous populations},
  author={Santos, Francisco C and Pacheco, Jorge M and Lenaerts, Tom},
  journal={Proceedings of the National Academy of Sciences},
  volume={103},
  number={9},
  pages={3490--3494},
  year={2006},
  publisher={National Academy of Sciences}
}

@article{wooldridge1995intelligent,
  title={Intelligent agents: Theory and practice},
  author={Wooldridge, Michael and Jennings, Nicholas R},
  journal={The knowledge engineering review},
  volume={10},
  number={2},
  pages={115--152},
  year={1995},
  publisher={Cambridge University Press}
}

@article{masuda2012evolution,
  title={Evolution of cooperation driven by zealots},
  author={Masuda, Naoki},
  journal={Scientific reports},
  volume={2},
  number={1},
  pages={646},
  year={2012},
  publisher={Nature Publishing Group UK London}
}

@article{cimpeanu2025evolutionary,
  title={The evolutionary advantage of guilt: co-evolution of social and non-social guilt in structured populations},
  author={Cimpeanu, Theodor and Pereira, Luis Moniz and others},
  journal={Journal of the Royal Society Interface},
  volume={22},
  number={228},
  year={2025},
  publisher={The Royal Society}
}

@article{bara2022enabling,
  title={Enabling imitation-based cooperation in dynamic social networks},
  author={Bara, Jacques and Turrini, Paolo and Andrighetto, Giulia},
  journal={Autonomous Agents and Multi-Agent Systems},
  volume={36},
  number={2},
  pages={34},
  year={2022},
  publisher={Springer}
}

@article{perera2026hybrid,
  title={Hybrid Human-Agent Social Dilemmas in Energy Markets},
  author={Perera, Isuri and de Nijs, Frits and Garcia, Julian},
  journal={arXiv preprint arXiv:2603.11834},
  year={2026}
}

@article{han2021or,
  title={When to (or not to) trust intelligent machines: Insights from an evolutionary game theory analysis of trust in repeated games},
  author={Han, The Anh and Perret, Cedric and Powers, Simon T},
  journal={Cognitive Systems Research},
  volume={68},
  pages={111--124},
  year={2021},
  publisher={Elsevier}
}

@article{song2026network,
  title={Network reciprocity turns cheap talk into a force for cooperation},
  author={Song, Zhao and Shen, Chen and Han, The Anh},
  journal={Journal of Theoretical Biology},
  volume={617},
  pages={112303},
  year={2026},
  publisher={Elsevier}
}

@article{hammond2025multi,
  title={Multi-agent risks from advanced ai},
  author={Hammond, Lewis and Chan, Alan and Clifton, Jesse and Hoelscher-Obermaier, Jason and Khan, Akbir and McLean, Euan and Smith, Chandler and Barfuss, Wolfram and Foerster, Jakob and Gaven{\v{c}}iak, Tom{\'a}{\v{s}} and others},
  journal={arXiv preprint arXiv:2502.14143},
  year={2025}
}

@article{kobis2025delegation,
  title={Delegation to artificial intelligence can increase dishonest behaviour},
  author={K{\"o}bis, Nils and Rahwan, Zoe and Rilla, Raluca and Supriyatno, Bramantyo Ibrahim and Bersch, Clara and Ajaj, Tamer and Bonnefon, Jean-Fran{\c{c}}ois and Rahwan, Iyad},
  journal={Nature},
  volume={646},
  number={8083},
  pages={126--134},
  year={2025},
  publisher={Nature Publishing Group UK London}
}

@article{lamarins2024importance,
  title={The importance of network spatial structure as a driver of eco-evolutionary dynamics},
  author={Lamarins, Ama{\"\i}a and Pr{\'e}vost, Etienne and Carlson, Stephanie M and Buoro, Mathieu},
  journal={Ecography},
  volume={2024},
  number={1},
  pages={e06933},
  year={2024},
  publisher={Wiley Online Library}
}

@article{galam2007role,
  title={The role of inflexible minorities in the breaking of democratic opinion dynamics},
  author={Galam, Serge and Jacobs, Frans},
  journal={Physica A: Statistical Mechanics and its Applications},
  volume={381},
  pages={366--376},
  year={2007},
  publisher={Elsevier}
}

@article{han2022voluntary,
  title={Voluntary safety commitments provide an escape from over-regulation in AI development},
  author={Han, The Anh and Lenaerts, Tom and Santos, Francisco C and Pereira, Luis Moniz},
  journal={Technology in Society},
  volume={68},
  pages={101843},
  year={2022},
  publisher={Elsevier}
}

@article{bago2020fake,
  title={Fake news, fast and slow: Deliberation reduces belief in false (but not true) news headlines.},
  author={Bago, Bence and Rand, David G and Pennycook, Gordon},
  journal={Journal of experimental psychology: general},
  volume={149},
  number={8},
  pages={1608},
  year={2020},
  publisher={American Psychological Association}
}

@article{masuda2015opinion,
  title={Opinion control in complex networks},
  author={Masuda, Naoki},
  journal={New Journal of Physics},
  volume={17},
  number={3},
  pages={033031},
  year={2015},
  publisher={IOP Publishing}
}

@article{nowak2006five,
  title={Five rules for the evolution of cooperation},
  author={Nowak, Martin A},
  journal={science},
  volume={314},
  number={5805},
  pages={1560--1563},
  year={2006},
  publisher={American Association for the Advancement of Science}
}

@article{qi2025facts,
  title={Facts about fact-checkers: Comparing credibility perceptions, usage, and sharing of different fact-checking sources},
  author={Qi, Li and Wang, Laurent H and Liu, Xingyu and Metzger, Miriam J},
  journal={Journal of Media Psychology},
  year={2025},
  publisher={Hogrefe Publishing}
}

@article{lewandowsky2017beyond,
  title={Beyond misinformation: Understanding and coping with the “post-truth” era},
  author={Lewandowsky, Stephan and Ecker, Ullrich KH and Cook, John},
  journal={Journal of applied research in memory and cognition},
  volume={6},
  number={4},
  pages={353--369},
  year={2017},
  publisher={Elsevier}
}

@article{becker2017network,
  title={Network dynamics of social influence in the wisdom of crowds},
  author={Becker, Joshua and Brackbill, Devon and Centola, Damon},
  journal={Proceedings of the national academy of sciences},
  volume={114},
  number={26},
  pages={E5070--E5076},
  year={2017},
  publisher={National Academy of Sciences}
}

@article{sulkin2001habermas,
  title={Habermas in the lab: A study of deliberation in an experimental setting},
  author={Sulkin, Tracy and Simon, Adam F},
  journal={Political Psychology},
  volume={22},
  number={4},
  pages={809--826},
  year={2001},
  publisher={Wiley Online Library}
}

@inproceedings{paiva2018engineering,
  title={Engineering pro-sociality with autonomous agents},
  author={Paiva, Ana and Santos, Fernando and Santos, Francisco},
  booktitle={Proceedings of the AAAI conference on artificial intelligence},
  volume={32},
  number={1},
  year={2018}
}

@article{mu2024multi,
  title={Multi-agent, human--agent and beyond: a survey on cooperation in social dilemmas},
  author={Mu, Chunjiang and Guo, Hao and Chen, Yang and Shen, Chen and Hu, Die and Hu, Shuyue and Wang, Zhen},
  journal={Neurocomputing},
  volume={610},
  pages={128514},
  year={2024},
  publisher={Elsevier}
}

@article{zimmaro2024emergence,
  title={Emergence of cooperation in the one-shot Prisoner’s dilemma through Discriminatory and Samaritan AIs},
  author={Zimmaro, Filippo and Miranda, Manuel and Fern{\'a}ndez, Jos{\'e} Mar{\'\i}a Ramos and Moreno L{\'o}pez, Jes{\'u}s A and Reddel, Max and Widler, Valeria and Antonioni, Alberto and others},
  journal={Journal of the Royal Society Interface},
  volume={21},
  number={218},
  year={2024},
  publisher={The Royal Society}
}

@article{sarkadi2024self,
  title={Self-governing hybrid societies and deception},
  author={Sarkadi, {\c{S}}tefan},
  journal={ACM Transactions on Autonomous and Adaptive Systems},
  volume={19},
  number={2},
  pages={1--24},
  year={2024},
  publisher={ACM New York, NY}
}

@article{castelfranchi2000artificial,
  title={Artificial liars: Why computers will (necessarily) deceive us and each other},
  author={Castelfranchi, Cristiano},
  journal={Ethics and Information Technology},
  volume={2},
  number={2},
  pages={113--119},
  year={2000},
  publisher={Springer}
}

@article{sarkadi2024deceptive,
  title={Deceptive AI and society},
  author={Sarkadi, {\c{S}}tefan},
  journal={IEEE Technology and society magazine},
  volume={42},
  number={4},
  pages={77--86},
  year={2024},
  publisher={IEEE}
}

@article{sarkadi2019modelling,
  title={Modelling deception using theory of mind in multi-agent systems},
  author={Sarkadi, {\c{S}}tefan and Panisson, Alison R and Bordini, Rafael H and McBurney, Peter and Parsons, Simon and Chapman, Martin},
  journal={AI Communications},
  volume={32},
  number={4},
  pages={287--302},
  year={2019},
  publisher={SAGE Publications Sage UK: London, England}
}

@article{sarkadi2021evolution,
  title={The evolution of deception},
  author={Sarkadi, {\c{S}}tefan and Rutherford, Alex and McBurney, Peter and Parsons, Simon and Rahwan, Iyad},
  journal={Royal Society open science},
  volume={8},
  number={9},
  pages={201032},
  year={2021}
}

@inproceedings{sarkadi2023arms,
  title={An arms race in theory-of-mind: Deception drives the emergence of higher-level theory-of-mind in agent societies},
  author={Sarkadi, {\c{S}}tefan},
  booktitle={2023 IEEE International Conference on Autonomic Computing and Self-Organizing Systems (ACSOS)},
  pages={1--10},
  year={2023},
  organization={IEEE}
}

@inproceedings{sarkadi2024triangles,
  title={The Triangles of Dishonesty: Modelling the Evolution of Lies, Bullshit, and Deception in Agent Societies},
  author={Sarkadi, {\c{S}}tefan and Lewis, Peter R.},
  booktitle={Proceedings of the 23rd International Conference on Autonomous Agents and Multiagent Systems (AAMAS 2024)},
  pages={1645--1653},
  year={2024},
  publisher={IFAAMAS}
}

@inproceedings{santos2019evolution,
  title={Evolution of collective fairness in hybrid populations of humans and agents},
  author={Santos, Fernando P and Pacheco, Jorge M and Paiva, Ana and Santos, Francisco C},
  booktitle={Proceedings of the AAAI conference on artificial intelligence},
  volume={33},
  number={01},
  pages={6146--6153},
  year={2019}
}

@article{alon2026impomdp,
  title={{$\aleph$}-IPOMDP: Mitigating deception in a cognitive hierarchy with off-policy counterfactual anomaly detection},
  author={Alon, Nitay and Barnby, Joseph M and Sarkadi, Stefan and Schulz, Lion and Rosenschein, Jeffrey S and Dayan, Peter},
  journal={Journal of Artificial Intelligence Research},
  volume={85},
  year={2026}
}

@article{szabo2007evolutionary,
  title={Evolutionary games on graphs},
  author={Szab{\'o}, Gy{\"o}rgy and Fath, Gabor},
  journal={Physics reports},
  volume={446},
  number={4-6},
  pages={97--216},
  year={2007},
  publisher={Elsevier}
}

@article{levine_truth-default_2014,
    title = {Truth-{Default} {Theory} ({TDT}): {A} {Theory} of {Human} {Deception} and {Deception} {Detection}},
    volume = {33},
    issn = {0261-927X},
    shorttitle = {Truth-{Default} {Theory} ({TDT})},
    url = {https://doi.org/10.1177/0261927X14535916},
    doi = {10.1177/0261927X14535916},
    language = {EN},
    number = {4},
    urldate = {2025-03-16},
    journal = {Journal of Language and Social Psychology},
    author = {Levine, Timothy R.},
    month = sep,
    year = {2014},
    note = {Publisher: SAGE Publications Inc},
    pages = {378--392},
}

%%%%%%%%%%%%%%%%%%%%%%%%%%%%%%%%%%%%%%%%%%%%%%%%%%%%%%%%%%%%%%%%%%%%%%%%

\end{document}